\documentclass{article}
\usepackage{dcase2021_arXiv,amsmath,graphicx,url,times,booktabs, tabularx}
\usepackage{cite}
\usepackage{subfigure}

\title{ToyADMOS2: Another dataset of miniature-machine operating sounds for anomalous sound detection under domain shift conditions}

\name{Noboru Harada, Daisuke Niizumi, Daiki Takeuchi, Yasunori Ohishi}
\secondlinename{Masahiro Yasuda and Shoichiro Saito}
\vspace{-5pt}
\address{\vspace{-5pt}
  NTT Corporation, Japan\\
}

\begin{document}

\ninept
\maketitle

\begin{sloppy}

\begin{abstract}
This paper proposes a new large-scale dataset called "ToyADMOS2" for anomaly detection in machine operating sounds (ADMOS). As did for our previous ToyADMOS dataset, we collected a large number of operating sounds of miniature machines (toys) under normal and anomaly conditions by deliberately damaging them but extended with providing controlled depth of damages in anomaly samples. Since typical application scenarios of ADMOS often require robust performance under domain-shift conditions, the ToyADMOS2 dataset is designed for evaluating systems under such conditions. The released dataset consists of two sub-datasets for machine-condition inspection: fault diagnosis of machines with geometrically fixed tasks and fault diagnosis of machines with moving tasks. Domain shifts are represented by introducing several differences in operating conditions, such as the use of the same machine type but with different machine models and parts configurations, different operating speeds, microphone arrangements, etc. Each sub-dataset contains over 27 k samples of normal machine-operating sounds and over 8 k samples of anomalous sounds recorded with five to eight microphones. The dataset is freely available for download at https://github.com/nttcslab/ToyADMOS2-dataset and https://doi.org/10.5281/zenodo.4580270.
\end{abstract}

\begin{keywords}
Anomaly detection in sounds, machine operating sounds, product inspection, dataset, domain shift conditions
\end{keywords}

\vspace{-1pt}
\section{Introduction}
\vspace{-5pt}
Recently, several research efforts have focused on anomaly detection. An anomaly detection task is designed to detect anomaly states by learning only normal condition data. Microphones have been used as sensors to detect anomalies, referred to as anomaly sound detection (ASD) or acoustic condition monitoring \cite{Koizumi_IEEE2019, Rushe_ICASSP2019, Koizumi_EUSIPCO2017, Kawachi_ICASSP2018, Koizumi_ICASSP2019, Kawachi_ICASSP2019, Koizumi_WASPAA2019, Koizumi_ICASSP2020_01, Heinicke_2015}. This task setting is different from other sound event detection tasks such as gunshot detection\cite{Valenzise2007}.

In general, it is very difficult or almost impossible to collect massive anomaly data. The ToyADMOS \cite{Koizumi_WASPAA2019_01} and MIMII  \cite{Purohit_DCASE2019_01} datasets are the first ones to be used for evaluating anomaly detection systems using sound. These datasets enable us to compare the performance of different systems. In 2020, a number of systems from various research organizations in academia and industry were submitted to ``DCASE 2020 Challenge Task 2, Unsupervised detection of anomalous sounds for machine condition monitoring \cite{Koizumi_DCASE2020}.''
The submitted systems performed quite well on the task, showing the great potential of applying deep-learning-based systems for unsupervised anomaly detection tasks \cite{Giri2020, Daniluk2020, Primus2020, Vinayavekhin2020, Hayashi2020, Zhou2020}.

However, in reality, the given application ASD scenario in the DCASE 2020 challenge was not ideal compared to realistic cases. The task setting was too basic, and the task requirements were much easier than it would be for typical ones in practical applications, where for example, the same machine type but different models are used at different operating speeds, and the conditions are not given as training data. Several independent research groups have tackled domain-shift or domain-adaptation related tasks \cite{Yamaguchi_ICASSP2019, QWang_IEEE2019, alma99117193593805503,  Kumagai2019TransferAD, Yang2020AnomalyDW, Vincent_Wannes_Jesse_2020}, but few open datasets that could serve this need have been made available. Though the previous ToyADMOS dataset has some data variations that can be used for testing domain-shift conditions, we would like to have more variations on the test configuration.

When evaluating the performance of ASD systems, the statistical characteristics of anomalous sound samples should be different from that of normal samples. However, if the difference is too significant, the anomaly detection task might not be difficult enough to evaluate system performance. One way of controlling the difficultly of the test configuration is to adjust the signal-to-noise ratio (SNR) of the added noise level, but noise reduction techniques, such as the ones in \cite{Kawagucgi_ICASSP2019, Daniluk2020}, can be used to mitigate the difficulty of the task. Therefore, there should be a way to control the difficulty, without relying on the SNR.

The difficulty of the task under domain shifts can be controlled by changing the statistical difference among normal samples across domains and/or the statistical difference between normal and anomaly sound samples within a domain. In designing challenging test conditions, it is nice to strike an appropriate balance between these two approaches.

To address the application scenarios discussed above, we provide yet another ADMOS dataset called ToyADMOS2. The ToyADMOS2 dataset adds more variations on condition arrangements dedicated to domain shifts. Like we did for the previous ToyADMOS dataset, we collected normal and anomalous operating sounds of miniature machines by deliberately damaging their components. 
The ToyADMOS2 dataset has the following characteristics:
\begin{itemize}
\item Designed for two ADMOS tasks: product inspection (toy car) and fault diagnosis for a moving machine (toy train).
\item Provides controlled domain-shift conditions on machine models, part configurations, operating speeds, microphone models and arrangements, and environmental noise.
\item Enables control of depth of damage in anomaly samples to provide choices on significance levels of statistical differences between normal and anomalous samples.
\end{itemize}
Note that the ToyADMOS2 dataset 
can be used along with the previous ToyADMOS dataset to provide a larger variety of test conditions. 

The proposed ToyADMOS2 dataset is freely available for download at https://github.com/nttcslab/ToyADMOS2-dataset, and https://doi.org/10.5281/zenodo.4580270. The license and explanation of some of its uses are also available at those links.

\begin{figure*}[t]
  \centering
  \vspace{-7mm}
  \includegraphics[width=0.95\textwidth]{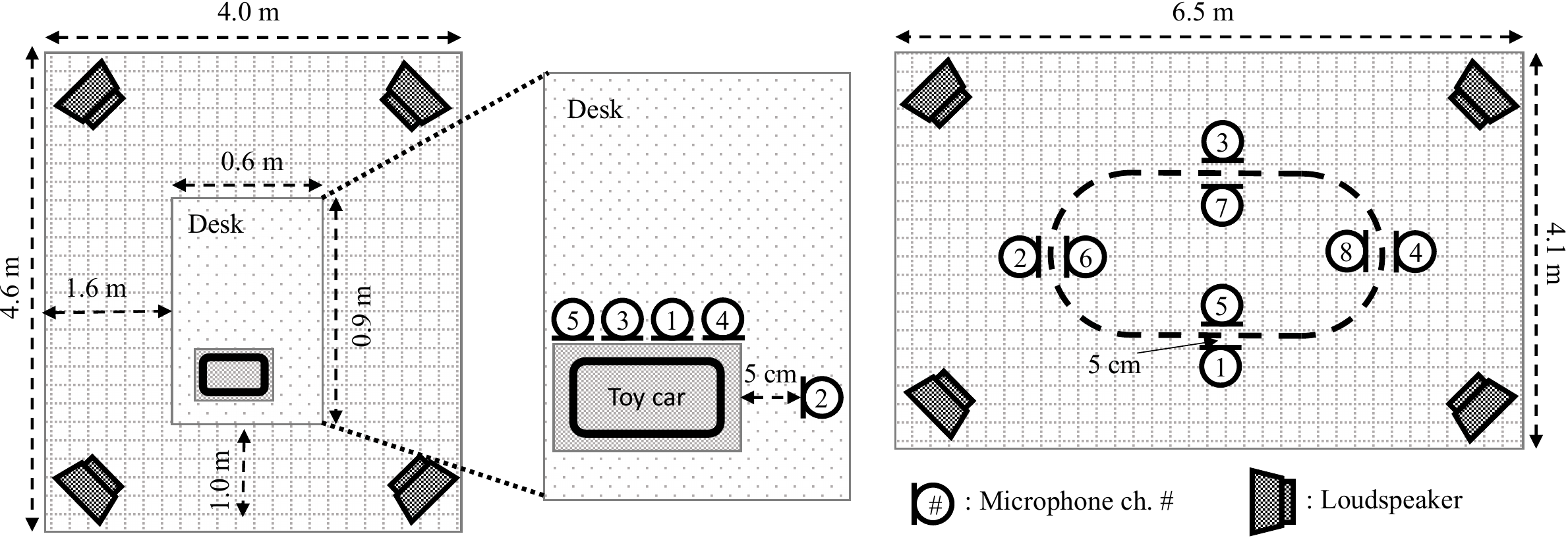}
  \\\vspace{-3mm}\subfigure[]{}\hspace{80mm}\subfigure[]{}\vspace{-3mm}
  \caption{Recording-room layouts and microphone arrangements.}
  \vspace{-4mm}
  \label{fig:imgs1}
\end{figure*}

\begin{figure}[tbh!]
  \centering
  \includegraphics[width=1.0\linewidth]{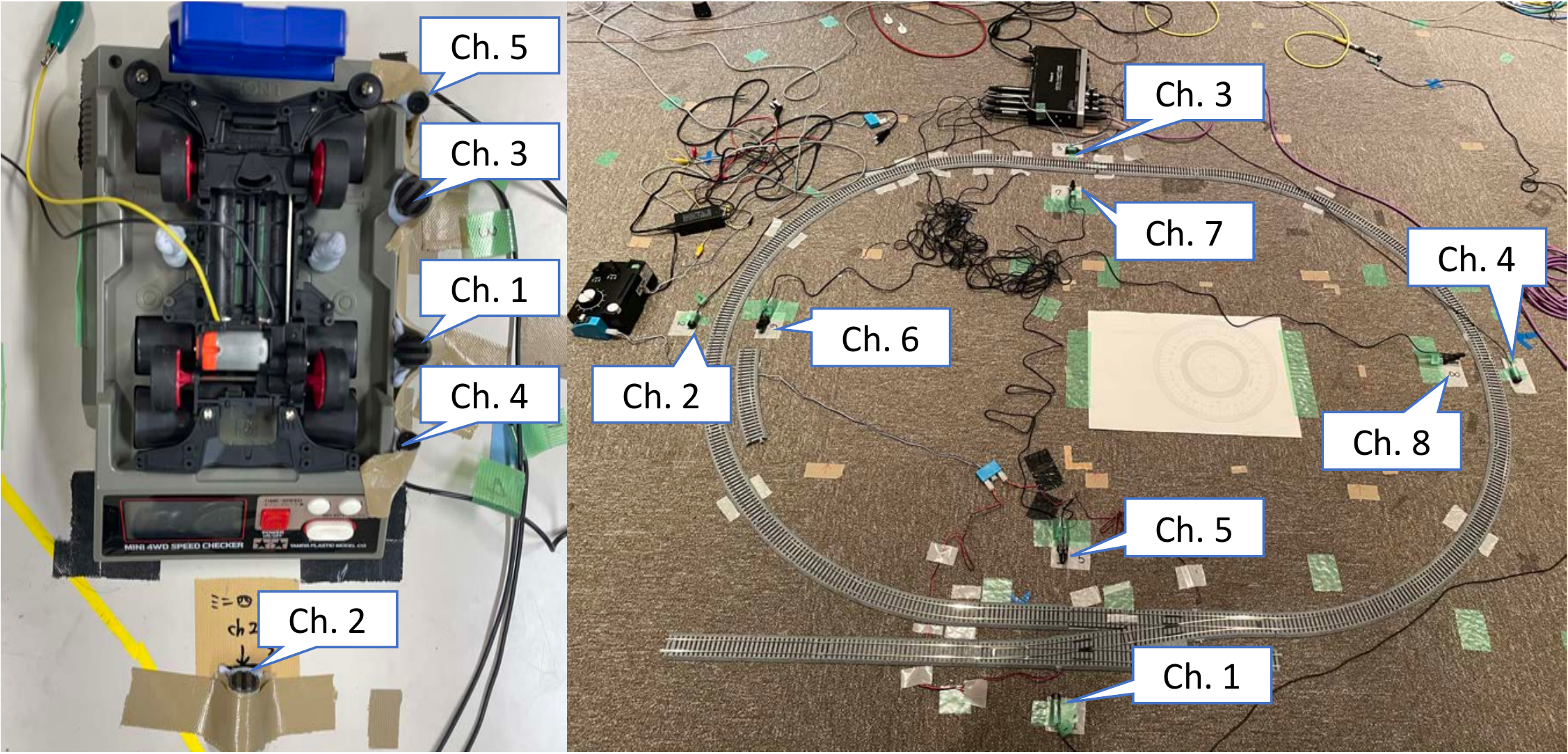}
  \\\vspace{-3mm}\hspace{-15mm}\subfigure[]{}\hspace{40mm}\subfigure[]{}\vspace{-4mm}
  \caption{Images of microphone arrangements.}
  \label{fig:imgs2}
\end{figure}

\begin{figure}[tbh!]
  \centering
  \includegraphics[width=1.08\linewidth]{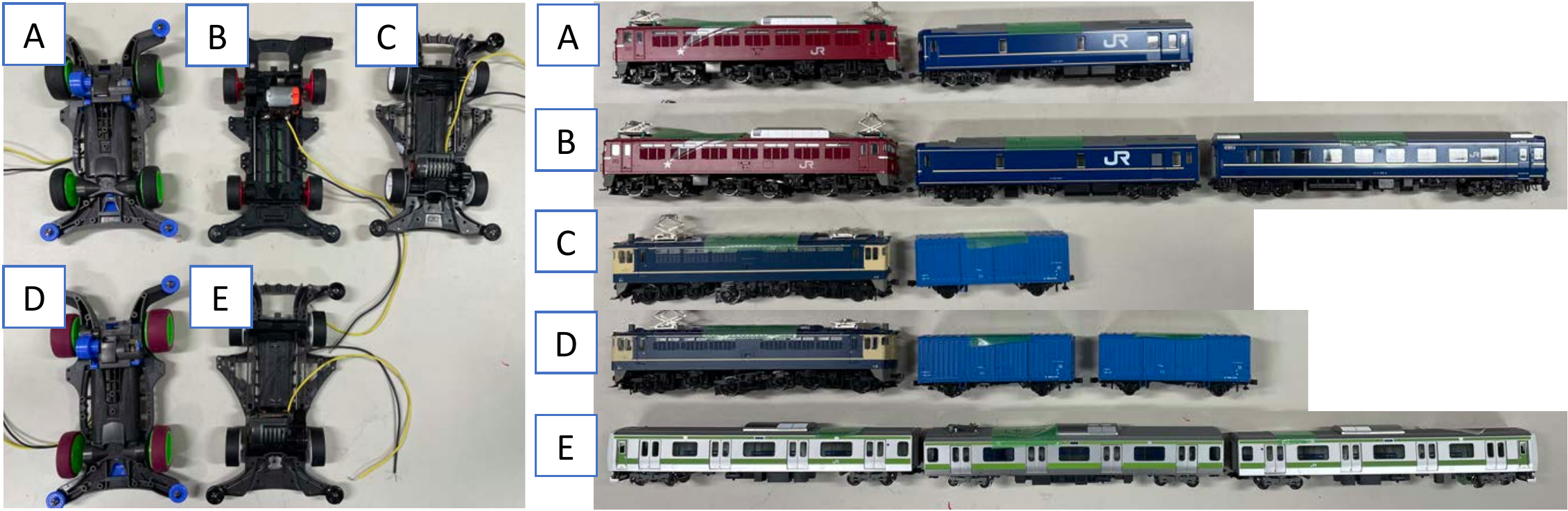}
  \\\vspace{-3mm}\hspace{-15mm}\subfigure[]{}\hspace{40mm}\subfigure[]{}\vspace{-4mm}
  \caption{Images of toy-model configurations.}
  \label{fig:imgs3}
\end{figure}


\section{Dataset overview}
\vspace{-4pt}

The ToyADMOS2 dataset consists of two sub-datasets for two types of ADMOS tasks. Each sub-dataset has three domain shift conditions.
A different type of toy, namely a toy car and a toy train, is used for each task.

\begin{description}
\item[Toy car:] Designed for the product-inspection task. The toy car runs on an inspection device. Sound data were collected with five microphones arranged close to the inspection device, as shown in Figs.~\ref{fig:imgs1}~(a) and \ref{fig:imgs2}~(a). The setting is similar to the toy car sub-dataset in ToyADMOS, but the machine models, part configurations, operating speed settings, and microphone arrangements are different. The toy-car models are shown in Fig.~\ref{fig:imgs3}~(a).
\item[Toy train:] Designed for fault diagnosis of a moving machine. The toy train runs on a railway track. Sound data were collected with eight microphones surrounding the track, as shown in Figs.~\ref{fig:imgs1}~(b) and \ref{fig:imgs2}~(b). The setting is similar to that of the toy train sub-dataset of the ToyADMOS dataset, but the machine models, cargo arrangements, operating speed settings, and microphone arrangements are different. The toy-train models are shown in Fig.~\ref{fig:imgs3}~(b).
\end{description}

To collect various normal and anomalous sounds depending on individual differences, operating sounds were recorded using several models with different part configurations.

\begin{description}
\item[Normal sound:] Operating sound when the target machine operates normally in accordance with its specifications.
\item[Anomalous sound:] Operating sound when the target machine was forced to operate anomalously by deliberately damaging its components or adding extraneous objects. The anomaly level was controlled by changing the depth of the damage deliberately made on the parts.
\item[Environmental noise:] Environmental noise for simulating a factory or other locations such as a roadside. Noise samples were collected at several locations with multiple microphones. These sounds were emitted from four loudspeakers positioned at the corners of each recording room and recorded with the same microphone configurations used for recording the normal and anomalous sound samples. In addition to newly recorded noise samples, noises from ToyADMOS dataset were also used.
\end{description}

Omnidirectional dynamic (SURE SM11-CN) and condenser microphones (TOMOCA EM-700), were used to collect these sounds. All sounds were stored as multiple WAV files.

In ToyADMOS2, several application scenarios representing domain-shift conditions were designed. Examples of possible domain-shift scenarios are as follows:

\begin{description}
\item[Model and parts configuration shift:] The normal and anomaly data were recorded using several different machine models and parts configurations.
\item[Operating speed shift:] Operating speed was varied. Normal and anomaly data were recorded under several different speed levels. 
\item[Microphone and environmental noise shift:] Different environmental noise was recorded at different positions using different types of microphones. 
\end{description}

\newpage

\section{Details of sub-datasets}

\subsection{Toy-car sub-dataset}
\vspace{-1mm}

We assumed a product inspection task and took up the task of detecting anomalous sounds from the running sound of a toy car on an inspection device, as shown in Fig.~\ref{fig:imgs2}~(a). The miniature car machine was a toy car called "mini 4WD", whose four wheels are driven by a small motor through gears and a shaft, was used. A stabilized power supply were connected to the motor, and running sounds on an inspection device were recorded with five microphones.
There were five machine configurations, A, B, C, D and E, as shown in Fig.~\ref{fig:imgs3}~(a). In each configuration, different machine chassis models and parts were used.

To control operating speed, five voltage levels, 2.8, 3.1, 3.4, 3.7, and 4.0 V, were provided through the stabilized power supply. Each recorded WAV file contains a normal or anomalous sound of a toy car. The duration of the normal and anomalous sound samples is 12 s. This results in over 35 k samples times five channels; in total over 177 k sound samples. Anomalous sounds were generated by deliberately damaging the shaft, gears, and tires. Three damage depth levels were provided to each part, as shown in Table~\ref{tab:tab1}. In total, over 8 k samples of anomalous sounds were recorded with their combinations (300 patterns) including three different depths of damage.

For recording environmental noise, four types of noise files were played through the four loudspeakers and recorded with the same microphone setting shown in Figs.~\ref{fig:imgs1}~(a) and \ref{fig:imgs2}~(a), and normal and anomalous sounds were recorded. Three dynamic and two condenser microphones were used. Some of the noise files were newly recorded; others were taken from ToyADMOS.

\vspace{-2mm}
\subsection{Toy-train sub-dataset}
\vspace{-1mm}

We assumed the use of fault diagnosis of a moving machine task of anomalous sounds from the running sound of a toy train. We used an HO-scale model train which is a precisely detailed miniature, as shown in Fig.~\ref{fig:imgs3}~(b). Sound data were collected with eight microphones surrounding positioned inside and outside the railway track perimeter (four microphones each). The microphones and loudspeakers in the recording room were arranged as shown in Figs.~\ref{fig:imgs1}~(b) and \ref{fig:imgs2}~(b). Dynamic microphones were used as the four outer microphones, channels 1 to 4. Condenser microphones were used as the four inner microphones, channels 5 to 8. 
Domain shift conditions in each task configuration are shown in Table~\ref{tab:tab1}.
There are five machine configurations, A, B, C, D and E.

To control operating speed, five levels of speed control, 5, 6, 7, 8 and 9, were used. Each recorded WAV file contains normal or anomalous sounds of a toy train. The duration of the normal and anomalous sound samples are 12 s each. This results in over 35 k samples times two mic types; in total over 71 k sound samples. Anomalous sounds were generated by deliberately damaging the carriage and railway track.  Three depth levels of the damage was provided to each part, as shown in Table~\ref{tab:tab1}. In total, over 8 k samples of anomalous sounds were recorded with their combinations (300 patterns).

For recording environmental noise, four types of noise files were played through the four loudspeakers and recorded with the same microphone setting shown in Figs.~\ref{fig:imgs1}~(b) and \ref{fig:imgs2}~(b), that normal and anomalous sounds were recorded with. Four dynamic microphones, channels 1 to 4, and four condenser microphones, channels 5 to 8, were used. Some of noise files were newly recorded; others were taken from ToyADMOS. 

\vspace{-4mm}
Further details of the ToyADMOS2 dataset are available at https://github.com/nttcslab/ToyADMOS2-dataset.


\begin{table}[t!]
  \vspace{-5mm}
  \caption{Anomaly conditions of sub-datasets}
  \label{tab:tab1}
  \centering
  \begin{tabular}{ l | l | l | l }
   \toprule
    \multicolumn{2}{c}{\textbf{Toy car}} & \multicolumn{2}{c}{\textbf{Toy train}} \\
    \midrule
    Parts & Condition & Parts & Condition\\
    \midrule
    Shaft   & - Bent        & Carriage & - Flat tire \\
    Gears   & - Deformed   &           & - Broken shaft  \\
                 & - Melted      &  Railway track & - Disjointed\\
    Wheels    &  - Damaged      &  (straight, curv) & - Obstructing stone\\
    \bottomrule
  \end{tabular}
\end{table}

\begin{table}[t!]
  \vspace{-4mm}
  \caption{Variation settings of sub-datasets}
  \label{tab:tab2}
  \centering
  \begin{tabular}{ l  c  c }
    \toprule
     & \textbf{Toy car} & \textbf{Toy train} \\
    \midrule
    Model variations  &  Five & Five\\
    Speed levels & Five & Five\\
    Mic. type and & Dynamic: 1-3 & Dynamic: 1-4\\
    ~~channel config. & Condenser: 4, 5 & Condenser: 5-8\\ 
    Noise type    &  \multicolumn{2}{c}{Four recordings}\\
    \midrule
    Normal samples & \multicolumn{2}{c}{1,094 samples$\times$5 models$\times$5 speed levels} \\ 
    ~~(Total hours) & (91 hours $\times$ 5 ch) & (91 hours$\times$2 ch-sets) \\
    Anomaly samples & \multicolumn{2}{c}{324 samples$\times$5 models$\times$5 speed levels}\\ 
    ~~(Total hours) & (27 hours$\times$5 ch) & (27 hours$\times$2 ch-sets) \\
    Noise samples & \multicolumn{2}{c}{24 hours per a channel}\\
    \bottomrule
  \end{tabular}
\end{table}

\begin{table}[t!]
  \vspace{-4mm}
  \caption{Example domain-shift task configurations}
  \label{tab:tab3}
  \centering
  \begin{tabular}{ l l c c }
    \toprule
    \textbf{Domain} & \multicolumn{2}{c}{\textbf{Configurations}} & \textbf{Training data}\\
    \midrule
     & Car/Train model & \hspace{-5mm}B & \\
    \textbf{Source} & Speed level & \hspace{-5mm}1, 3 & 1,500 normal\\
    \textbf{domain} & Mic. type &  \hspace{-5mm}Dynamic & samples \\
    & Noise &  \hspace{-5mm}N1 & \\
    \midrule
    \textbf{Target} & \textbf{Car/Train model} &  \hspace{-5mm}\textbf{D} & \\
    \textbf{domain 1:} & Speed level &   \hspace{-5mm}1, 3 & Four normal\\
    Model and & Mic. type &  \hspace{-5mm}Dynamic & samples\\
    parts shift & Noise &  \hspace{-5mm}N1 & \\
    \midrule
    \textbf{Target} & Car/Train model &  \hspace{-5mm}B & \\
    \textbf{domain 2:} & \textbf{Speed level} &  \hspace{-5mm}\textbf{2, 5} & Four normal\\
    Operating & Mic. type &  \hspace{-5mm}Dynamic & samples \\
    speed shift & Noise &  \hspace{-5mm}N1 & \\
    \midrule
    \textbf{Target}  & Car/Train model &  \hspace{-5mm}B & \\
    \textbf{domain 3:} & Speed level &  \hspace{-5mm}1, 3 & Four normal \\
    Mic. and & \textbf{Mic. type} &  \hspace{-5mm}\textbf{Condenser} & samples \\
    noise shift & \textbf{Noise} &  \hspace{-5mm}\textbf{N2} &\\
    \midrule
    \textbf{Target}  & \textbf{Car/Train model} &  \hspace{-5mm}\textbf{D} & \\
    \textbf{domain 4:} & \textbf{Speed level} &  \hspace{-5mm}\textbf{2, 5} & Four normal \\
    All of & \textbf{Mic. type} &  \hspace{-5mm}\textbf{Condenser} & samples \\
    above          & \textbf{Noise} &  \hspace{-5mm}\textbf{N2} & \\
    \bottomrule
    \multicolumn{4}{l}{Test data for each domain consists of 100 normal and 100}\\
    \multicolumn{4}{l}{ anomaly samples.}    \vspace{-2mm}
\vspace{-4mm}
  \end{tabular}
\end{table}



\section{Sample domain-shift task settings and benchmark}
\vspace{-2mm}
    
To show how to use the ToyADMOS2 dataset, we designed examples of task configurations for testing ASD systems under domain-shift conditions. Domain-shift configurations are shown in Table~\ref{tab:tab3}.


\begin{table*}[t!]
  \vspace{-10mm}
\caption{Benchmark AUC results of the DCASE 2020 Challenge task 2 baseline under domain-shift conditions}
  \label{tab:tab5}
  \centering
  \begin{tabular}{ l  c  c  c  c  c  c  c  c  c}
    \toprule
     & & \multicolumn{4}{c}{\textbf{Toy car}} & \multicolumn{4}{c}{\textbf{Toy train}} \\
    \midrule
     & Damage level & \textbf{Clean} & \textbf{6 dB} & \textbf{0 dB} & \textbf{-6 dB} & \textbf{Clean} &  \textbf{6 dB} & \textbf{0 dB} & \textbf{-6 dB} \\
    \midrule
     \textbf{Source domain} & high & \textbf{0.99} & 0.99 & 0.96 & 0.85 & \textbf{1.00} & 0.86 & 0.66 & 0.61 \\
     & mid & \textbf{0.99} & 0.95 & 0.94 & 0.80 & \textbf{0.98} & 0.66 & 0.57 & 0.56 \\
     & low & \textbf{1.00} & 0.96 & 0.95 & 0.80 & \textbf{0.83} & 0.57 & 0.59 & 0.49 \\
    & avg. & \textbf{0.99} & 0.97 & 0.95 & 0.82 & \textbf{0.94} & 0.70 & 0.61 & 0.55 \\
    \midrule
    \textbf{Target domain 1:} & high & 0.37 & 0.29 & 0.34 & 0.38 & 0.82 & 0.75 & 0.71 & 0.62 \\
    Model and parts shift & mid & 0.92 & 0.55 & 0.58 & 0.59 & 0.71 & 0.61 & 0.61 & 0.53 \\
     & low & 0.80 & 0.69 & 0.68 & 0.69 & 0.57 & 0.56 & 0.58 & 0.58 \\
     & avg. & 0.70 & 0.51 & 0.53 & 0.55 & 0.70 & 0.64 & 0.63 & 0.57 \\
    \midrule
    \textbf{Target domain 2:} & high &  0.77 & 0.78 & 0.83 & 0.76 & 0.71 & 0.65 & 0.62 & 0.57 \\
    Operating speed shift & mid & 0.61 & 0.63 & 0.65 & 0.66 & 0.65 & 0.61 & 0.51 & 0.52 \\
     & low & 0.64 & 0.72 & 0.67 & 0.69 & 0.52 & 0.52 & 0.48 & 0.44 \\
     & avg. & 0.68 & 0.71 & 0.71 & 0.70 & 0.63 & 0.59 & 0.54 & 0.51 \\
    \midrule
    \textbf{Target domain 3:} & high &0.55 & 0.53 & 0.45 & 0.52 & 0.72 & 0.62 & 0.57 & 0.57 \\
    Mic. type and noise shift & mid & 0.39 & 0.55 & 0.47 & 0.48 & 0.66 & 0.55 & 0.54 & 0.52 \\
    & low & 0.45 & 0.57 & 0.56 & 0.52 & 0.65 & 0.52 & 0.52 & 0.52 \\
    & avg. & 0.46 & 0.55 & 0.49 & 0.50 & 0.68 & 0.56 & 0.54 & 0.54 \\
    \midrule
    \textbf{Target domain 4:} & high & 0.55 & 0.71 & 0.59 & 0.65 & 0.60 & 0.57 & 0.53 & 0.49 \\
    All above shifts & mid & 0.44 & 0.64 & 0.54 & 0.56 & 0.51 & 0.45 & 0.44 & 0.45 \\
    & low & 0.52 & 0.71 & 0.55 & 0.65 & 0.51 & 0.45 & 0.42 & 0.44 \\
    & avg. & 0.50 & 0.69 & 0.56 & 0.62 & 0.54 & 0.49 & 0.46 & 0.46 \\
    \bottomrule
\vspace{-10mm}
  \end{tabular}
\end{table*}



We assume a typical application scenario with a task setting similar to that in \cite{QWang_IEEE2019}. A relatively large number of normal samples recorded in a source domain (e.g., 1,500 samples) are available as a training dataset; however, a very limited number of normal samples recorded in target domains (four samples each) are given for training. As shown in Table~\ref{tab:tab3}, there are four example domain-shift configurations: \textbf{Target domain 1}, Model and parts shift; \textbf{Target domain 2}, Operating speed shift; \textbf{Target domain 3}, Mic. type and noise shift; and \textbf{Target domain 4}, A mixture of shifts 1 to 3. 

For the source domain training dataset, 1,500 normal samples of Toy-car model B and Toy-train model B were used. For Toy car, normal sample data recorded with a dynamic microphone, ch 1, were used. For Toy train, normal sample data recorded by dynamic microphones, ch 1 to 4, were used. In those normal samples, operating speed was set to levels 1 and 3.
For target domains 1 to 4, only four normal samples were given as training data for each of target domains. 
Environmental noises N1 and N2 recorded by using microphones at the same positions as for the normal samples were used. SNRs, calculated with
\vspace{-2.1mm}
\begin{equation}
\label{eq1}
SNR = 20 \log_{10}\Big\{ \frac{1}{J} \sum{rms(S_j)} \Big\} / rms(N_k)
\vspace{-1.2mm}
\end{equation}
where $S_j$ is a recorded machine-operating sound sample and $N_k$ is a recorded environmental noise sample with the same duration of $S_j$, were set to $+\infty$ dB (clean), 6 dB, 0 dB, and -6 dB. After mixed with the noise, all sound samples were down-sampled at a sampling rate of 16 kHz.
Detailed conditions of each target domain are shown in Table~\ref{tab:tab3}.

To simplify the sample system, a baseline system of DCASE 2020 challenge task 2 \cite{Koizumi_DCASE2020} -- a simple unsupervised-ASD one with an auto encoder as a normal model -- was tested under the task setting described above.
All the training data -- 1,500 normal samples of the source domain, and other normal samples from target domains (four samples each) -- were merged to formulate a training dataset that contains 1,516 normal samples. The baseline system was trained for each SNR condition of Toy car and Toy train.

The trained baseline systems were tested using 200 unknown test samples that were not used for training. 
We calculated anomaly scores on each time frame for all the test WAV files as described in \cite{Koizumi_DCASE2020}.
Calculated area under curb (AUC) results are shown in Table~\ref{tab:tab5}. 
The results in bold show that the damage level is higher, anomaly detection is easier for the source domain/clean for Toy train. For Toy car, the AUC results were around 0.99 for all the levels.
Distribution of the anomaly scores for the source domain/clean samples are shown in Fig.~\ref{fig:imgs4}.


\begin{figure}[tbh!]
\vspace{-3mm}
  \centering
  \includegraphics[width=1.0\linewidth]{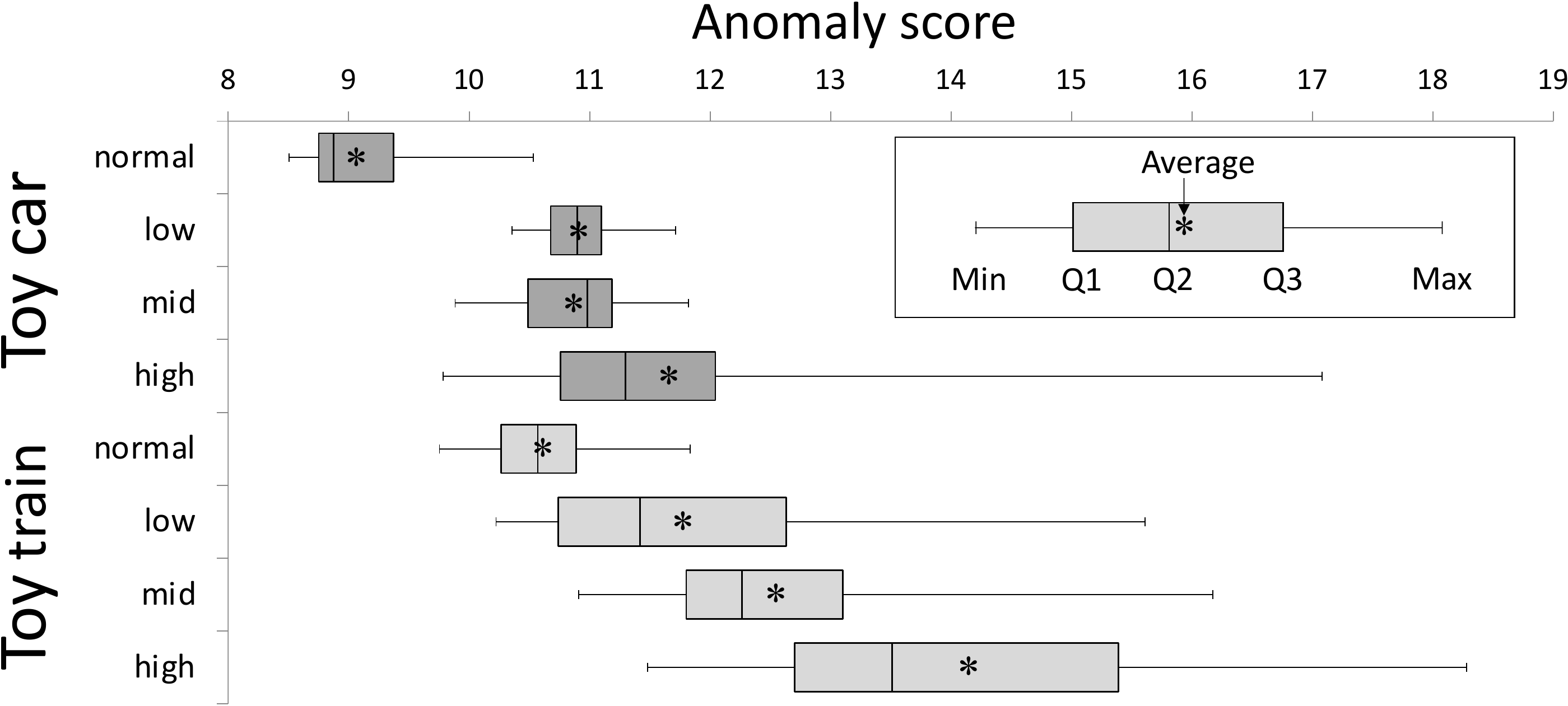}
  \vspace{-6mm}
  \caption{Anomaly scores of the source domain/clean samples.}
  \label{fig:imgs4}
\end{figure}

\vspace{-4mm}
\section{Conclusions}
\vspace{-2mm}

This paper proposed a new large-scale dataset called "ToyADMOS2" for ADMOS. The ToyADMOS2 dataset is designed for evaluating systems under domain-shift conditions and available for free dowonload. It consists of two sub-datasets for machine-condition inspection: fault diagnosis of machines with geometrically fixed tasks and fault diagnosis of machines with moving tasks. Domain shifts are represented by introducing several differences in operating conditions, such as the use of the same machine type but with different machine models and part configurations, different operating speeds, microphone arrangements, etc. Each sub-dataset contains over 27 k samples of normal machine-operating sounds and over 8 k samples of anomalous sounds recorded at a 48-kHz sampling rate. 
A subset of the ToyADMOS2 dataset was used in the DCASE 2021 challenge task 2: Unsupervised anomalous sound detection for machine condition monitoring under domain shifted conditions.


\newpage

\bibliographystyle{IEEEtran}
\bibliography{mybib}

%
%
%
%
%
%
%
%
%

\end{sloppy}
\end{document}